# Deep Kernel Gaussian Process Based Financial Market Predictions

Yong Shi[1, 2, 3], Wei Dai[1, 2, 3], Wen Long[1, 2, 3, *], Bo Li[1, 2, 3]

1 *School of Economics and Management, University of Chinese Academy of Sciences, No. 80 of Zhongguancun East Street, Haidian District, Beijing, 100190, P.R.China*

2 *Research Center on Fictitious Economy and Data Science, Chinese Academy of Sciences, No. 80 of Zhongguancun East Street, Haidian District, Beijing, 100190, P.R.China*

3 *Key Laboratory of Big Data Mining and Knowledge Management, Chinese Academy of Sciences , No. 80 of Zhongguancun East Street, Haidian District, Beijing, 100190, P.R.China*

* Corresponding author information:

E-mail: longwen@ucas.ac.cn;

Address: No. 80 of Zhongguancun East Street, Haidian District, Beijing, 100190, P.R.China

**Abstract**

The Gaussian Process with a deep kernel is an extension of the classic GP regression model and this extended model usually constructs a new kernel function by deploying deep learning techniques like long short-term memory networks. A Gaussian Process with the kernel learned by LSTM, abbreviated as GP-LSTM, has the advantage of capturing the complex dependency of financial sequential data, while retaining the ability of probabilistic inference. However, the deep kernel Gaussian Process has not been applied to forecast the conditional returns and volatility in financial market to the best of our knowledge. In this paper, grid search algorithm, used for performing hyper-parameter optimization, is integrated with GP-LSTM to predict both the conditional mean and volatility of stock returns, which are then combined together to calculate the conditional Sharpe Ratio for constructing a long-short portfolio. The experiments are performed on a dataset covering all constituents of Shenzhen Stock Exchange Component Index. Based on empirical results, we find that the GP-LSTM model can provide more accurate forecasts in stock returns and volatility, which are jointly evaluated by the performance of constructed portfolios. Further sub-period analysis of the experiment results indicates that the superiority of GP-LSTM model over the benchmark models stems from better performance in highly volatile periods.

# Introduction

It is widely accepted that stock price movement can be predicted to some extent and time series analysis has been a key instrument of portfolio construction and asset allocation. Empirical research conducted by Marquering and Verbeek (2004) has confirmed the joint economic value of predicting stock market returns as well as volatility. Autoregressive conditional heteroskedasticity (ARCH) model proposed by Engle(1982), Generalized Autoregressive Conditional Heteroskedasticity (GARCH) model proposed by Bollerslev(1986) and relevant extended models allow us to simultaneously describe the process of the conditional mean and volatility for stock returns. However, financial time series data has a significant time-varying characteristic (Giles et al., 2001) which makes the forecasting task a challenging problem.

In recent years, machine learning methods have been widely applied to image identification and natural language processing tasks for looser model assumption and better generalization ability compared with traditional statistical models. More and more researchers are in favor of deploying machine learning methods to solve financial prediction problems. Compared with the neural networks and the support vector machine, Gauss process (GP) has the advantage of providing probabilistic output. However, the classic GP model has a fixed-form function which limits the fitting ability of this model, so we are motivated to utilize a GP-LSTM model (Al-Shedivat et al., 2017), which extends the basic kernel function of GP with long-short term memory (LSTM) networks.

In this study, we combine the GP-LSTM model with grid search (GS) algorithm to forecast the conditional returns and volatility in financial market. The corresponding trading strategy is designed to prove the effectiveness of the method. The contributions of this paper mainly include the following points:

(1) GP-LSTM model has been applied to various fields because this model utilizes the strong fitting ability of LSTM networks while retaining statistical inference ability. To the best of our knowledge, GP-LSTM has not been applied to simultaneously forecast the conditional returns and volatility in financial market.

(2) By combining GP-LSTM model with GS algorithm, this paper minimizes the impact

caused by the manual tuning and improve the model performance.

(3) This paper applies this GP-LSTM method to the empirical study covering all constituents of Shenzhen Stock Exchange Component Index (SZSE COMP) and designs the corresponding trading strategy. In addition, Sharpe Ratio of the constructed portfolios is introduced to jointly measure the forecast performance of conditional return and volatility.

The rest of this paper is organized as follows. In second section, we review the related work concentrating on forecasting stock returns and volatility. The third section provides a detailed instruction of the GP-LSTM method, the benchmark models and the process of data preparation. The corresponding experiment results are presented in the fourth section. In the fifth section, further sub-period analysis is conducted to explore the source of GP-LSTM's superiority. The sixth section concludes this paper and points out the possible direction of future research.

## Related Work

### ARCH and GARCH family models

ARCH and GARCH family models are classic methods for forecasting financial returns and volatility. To describe the volatility clustering effect of financial asset price, ARCH model assumes that the conditional variance depends on the lagged squared error terms in the volatility equation. However, ARCH family models often need to set a high lag order, which greatly increases the number of parameters required to estimate. GARCH model is correspondingly proposed to solve this problem. As set in the GARCH model, variance of conditional returns depends not only on the lagged squared error terms, but also on lagged terms of the variance itself. In this way, the GARCH models with low lag order can be used instead of the ARCH models with high lag order.

In order to reflect the asymmetry impact of error terms in the volatility equation of GARCH and Nelson (1991), Glosten and Jagannathan (1992) extended the standard GARCH model to EARCH model and GJR-GARCH model respectively. In GARCH family models, Gaussian distribution is generally used to characterize the error term in mean equation, but the distribution of stock returns usually exhibits the characteristic of high peak and fat tail in realistic situations. Therefore, researchers made a natural extension for the error term from Gaussian to non-Gaussian (e.g. *t* distribution, *skewed t* distribution) for fitting financial data better.

**Machine learning methods**

*Artificial Neural Networks*

Artificial Neural Networks (ANN) consists of a set of units called artificial neurons, which are usually organized in several layers. According to Universal Approximation Theorem (Hornik et al., 1989), feedforward neural networks can approximate a *Borel* measurable function to any desired degree of accuracy if sufficiently many hidden units with arbitrary squashing functions are provided.

Donaldson (1997) constructed a nonlinear GARCH model based on ANN to predict stock index volatility. Hyup (2007) utilized GARCH model to generate input variables of an ANN to enhance the prediction ability in forecasting volatility. These studies empirically prove that ANN has practical value in the prediction of volatility.

*Support Vector Regression*

The support vector machine (SVM) proposed by Vapnik (1999) is also one of the most important machine learning models. In classification tasks, SVM classifies the data points by finding a hyperplane that has the maximum margin, i.e. the maximum distance between support vectors. As the linear separability may not be satisfied, the kernel function is introduced to map the data points in raw representation into a higher even infinite dimensional space. The support vector regression (SVR) is an extension of SVM model and transforms the discrete output of SVM into an ordinary 1-dimensional numerical space for regression tasks**.** Pérez-Cruz et al. (2003) utilized SVM algorithm to estimate the parameters of GARCH model, improving the forecasting accuracy of volatility. Peng (2018) also chose to estimate the mean and volatility equations of GARCH, Egarch and GJR-GARCH models respectively using support vector regression for the data of three cryptocurrencies and three currency exchange rates, which achieved more accurate predictions than the basic GARCH family models.

However, the studies for volatility prediction utilizing ANN or SVR usually choose a proxy variable as an approximation for the real volatility (variance of the conditional return). Hence, classic machine learning methods can not be directly applied in the prediction of real volatility.

*Brief introduction of Gaussian Process*

Similar to SVR model, kernel trick is also widely used in GP models. The covariance between each pair of data points, is measured in the kernel space and Bayesian theory is applied to predict the posterior distribution of out-of-sample data points. Compared with ANN and SVR models, the GP model has the advantages of providing probabilistic output and non-parametric inference flexibility. Wu (2014) constructed the Gaussian Process Volatility model based on the standard Gaussian Process and found that it outperformed the classic GARCH models.

However, in the previous applications of SVR and GP model, researchers often choose fixed-form kernel functions, which limits the fitting ability of these models. Al-Shedivat et al. (2017) proposed a new GP model in which the kernel can be learned by neural networks like LSTM networks and this new model has been successfully applied to many problems such as wind power prediction and automatic driving system. In realistic situations, the stocks always differ in size, industry and liquidity, leading to different modes of stock price movement. Therefore, it's natural to assume that hyper parameters which decide the signal-to-noise ratio and volatile degree should vary widely when applying GP model to different stocks. To solve this problem, GS algorithm can be used to find a good set of initial hyper parameters and reduce the impact of manual tuning.

In this paper, we combine the deep kernel GP model with grid search algorithm for the prediction problem of stock conditional mean and volatility, and construct the portfolio from constituent stocks of the SZSE COMP according to Sharpe Ratio.

# GP-LSTM model

## Model construction

## GP model

A GP model could be viewed as a generation process of the Gaussian distribution, and arbitrary finite variables yielded by a Gaussian Process have a joint Gaussian distribution which is determined by a mean function $m(X)$ such that $(m(X))_i = m(x_i)$ and a covariance kernel function $K(X,X)$ such that $(K(X,X))_{i,j} = k(x_i, x_j)$:

$$[f(x_1), f(x_2), ... f(x_n)] \sim N(m(X), K_{X,X}) \tag{1}$$

where $f$ represents a Gaussian Process, $x_i$ is a feature vector and $X = (x_1, x_2, ...., x_n)^T$. The covariance between each pair of data points is determined by the feature vectors $x_i, x_j$ and the kernel function $k$. Given the training set $(Y, X)$ and out-of-sample points $X_*$, we can predict conditional return and variance of $Y_*$ based on Bayesian method by the following formulas:

$$\begin{aligned} Y_* \mid X_*, X, Y, \sigma^2 &\sim N(E[Y_*], Cov[Y_*]) \\ E[Y_*] &= \mu_{X_*} + K_{X_*,X}[K_{X,X} + \sigma^2 I]^{-1} Y, \\ Cov[Y_*] &= K_{X_*,X_*} - K_{X_*,X}[K_{X,X} + \sigma^2 I]^{-1} K_{X,X} \end{aligned} \tag{2}$$

As shown in Fig. 1, we denote the feature vector and output variable at time $t$ as $x_t$, $y_t$ respectively in this paper.

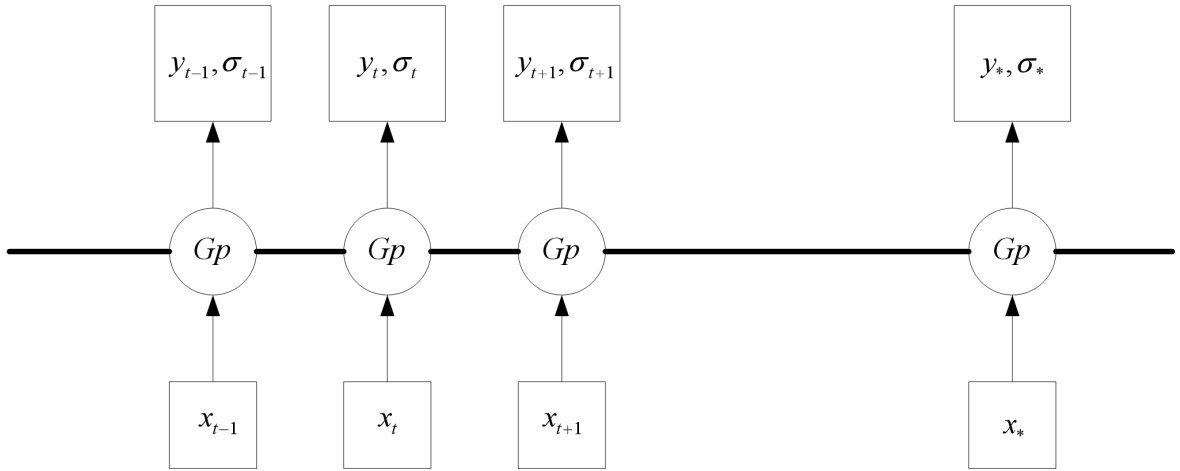

**Fig. 1** classic Gaussian Process model

*Learning the deep kernel*

The GP-LSTM model deployed in this paper utilizes a single-layer LSTM with 8 neurons to generate an advanced feature vector containing temporal information as the input of basic kernel function (see Eq. (3)).

$$x_t^{LSTM} = \phi((x_{t-1}, x_{t-2}, ..., x_{t-l})) = LSTM((x_{t-1}, x_{t-2}, ..., x_{t-l})) \tag{3}$$

Where $(x_{t-1}, x_{t-2}, ..., x_{t-l})$ represents the sequential input at time $t$ with the length of $l$. Hence, we can construct the deep kernel $K^{Deep}$ in Eq. (4).

$$K^{Deep} = K(X^{LSTM}, X^{LSTM}), (X^{LSTM})_t = x_t^{LSTM} \tag{4}$$

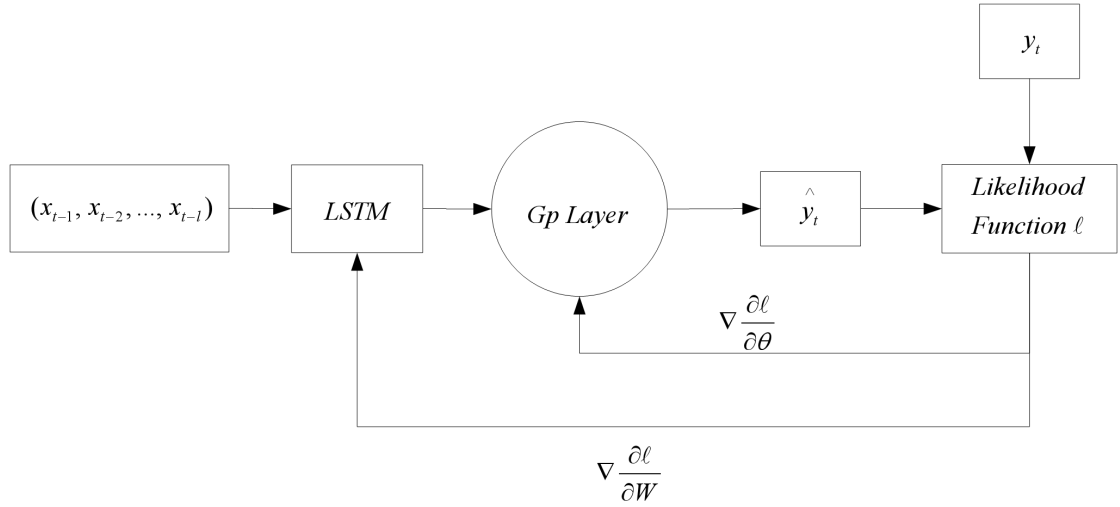

**Fig. 2** gradient optimization

We choose the Radial Basis Function ($RBF$) as the basic kernel function here, the form of $K_{ij}$ can be specified as:

$$K_{ij} = k(x_i, x_j) = \sigma_f^2 \exp(-1/2l^2 (x_i - x_j)^2) + \sigma_n^2 \delta_{ij} \tag{5}$$

As depicted in Fig.2, the hyperparameters $\theta$ of the GP layer and the parameters $W$ of the LSTM network will be trained jointly.

*Grid Search*

In the application of GP-LSTM model, squared-exponential covariance function is chosen as the kernel for the GP layer, which has three hyper parameters: length scale *L*, the signal variance $\sigma_f^2$ and the noise variance $\sigma_n^2$. Rasmussen et al. (2005) has illustrated that a Gaussian Process with different hyper-parameter combinations may all allow corresponding model residuals seem like white noise. It means that finding a good set of hyperparameters can be difficult for the $RBF$ kernel. As the hyper-parameter space is infinite, we utilize the heuristic grid search method to choose the optimal combination of initial hyper parameters according to the performance on validation set. Although the hyperparameters are optimized during the training process, choosing proper initial values can still improve the prediction performance. By assembling the above modules together, we utilize the GP-LSTM method (Fig. 3(a), Fig. 3(b)) to capture the price movement patterns of different stocks and reduce the manual tuning effect.

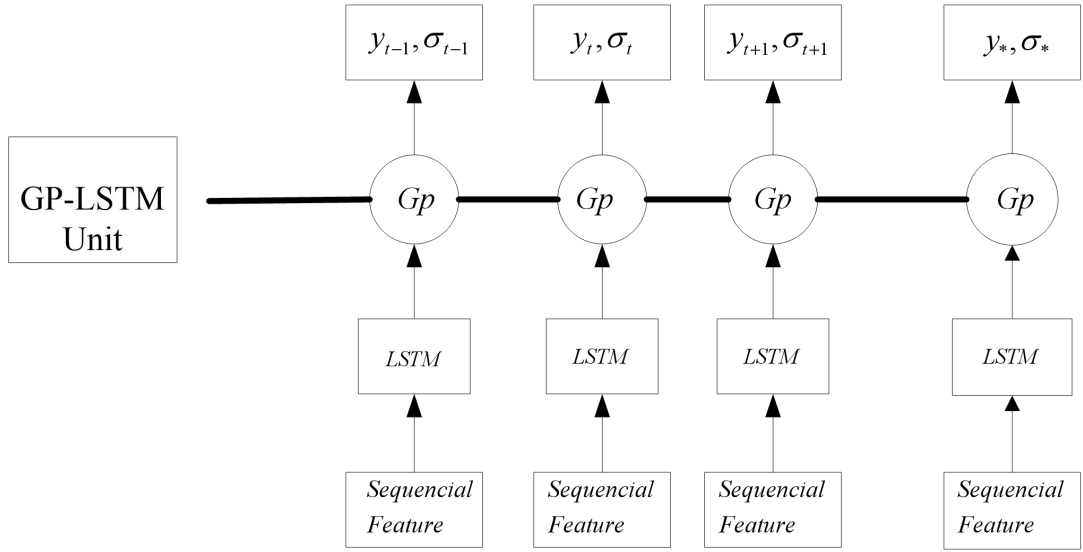

**Fig. 3(a)** GP-LSTM unit

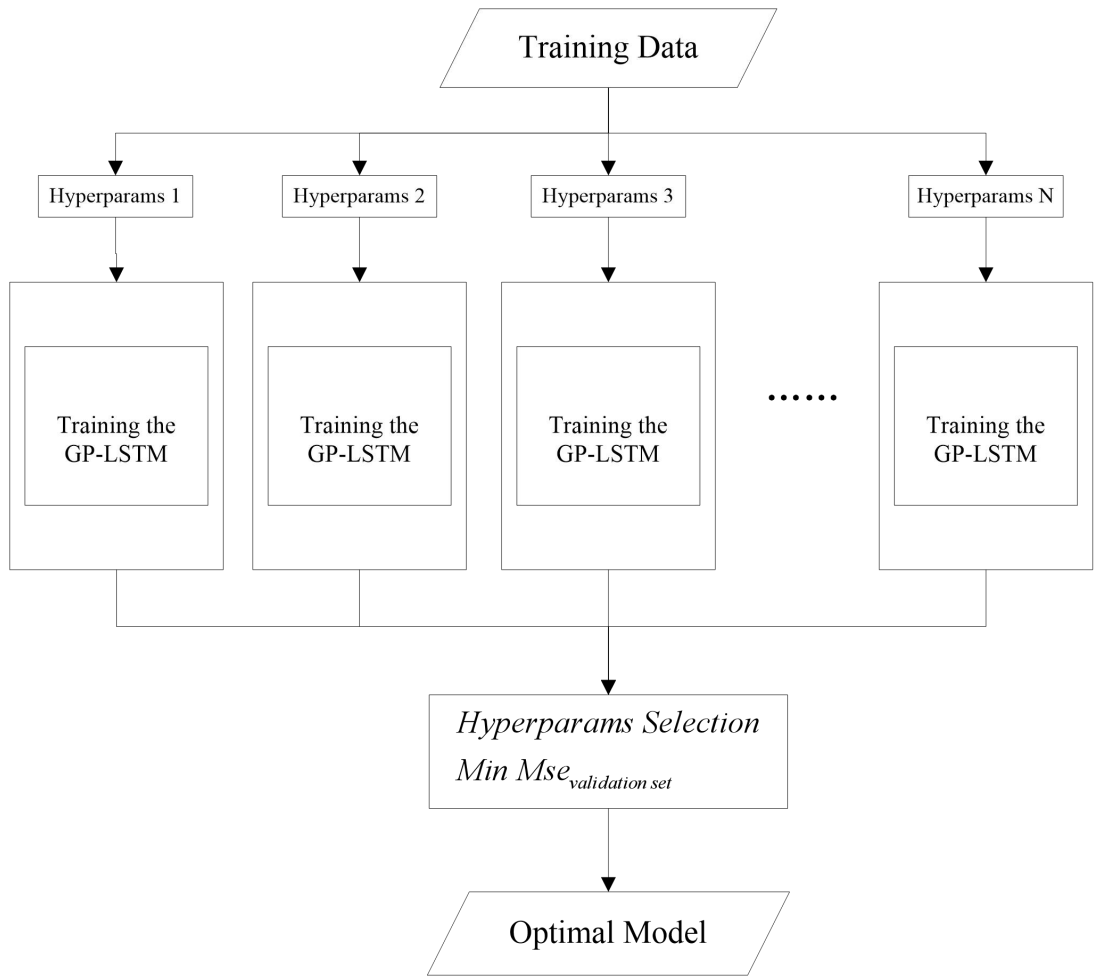

**Fig. 3(b)** GP-LSTM model combined with grid search algorithm

*Training process*

As described above, GP-LSTM unit (see Fig.3 (a)) can be divided into two parts: the classical GP regression model and the LSTM network which maps the raw representation of sequential data into advanced feature vector for the kernel function. For the LSTM part, we adopt fixed hyper parameters including learning rate, maximum learning epochs, batch size etc. For the GP part, we go through all the listed hyper-parameter combinations (Hyper parameters 1 – Hyper parameters N) to select the optimal one according to the Mean Squared Error ( *Mse* ) on the validation set.

Besides the selection of hyper-parameter combination, the remaining parameters of the kernel function $K$ and the LSTM networks need to be trained for fitting and prediction. Despite the strong ability to fit highly complex function, deep neural networks may easily meet with over-fitting problem. We can utilize early-stopping method to solve this problem. As shown in Fig.4, the early-stopping requires to evaluate model performance on validation set after each training cycle and the training process will be stopped when the generalization error begins to increase.

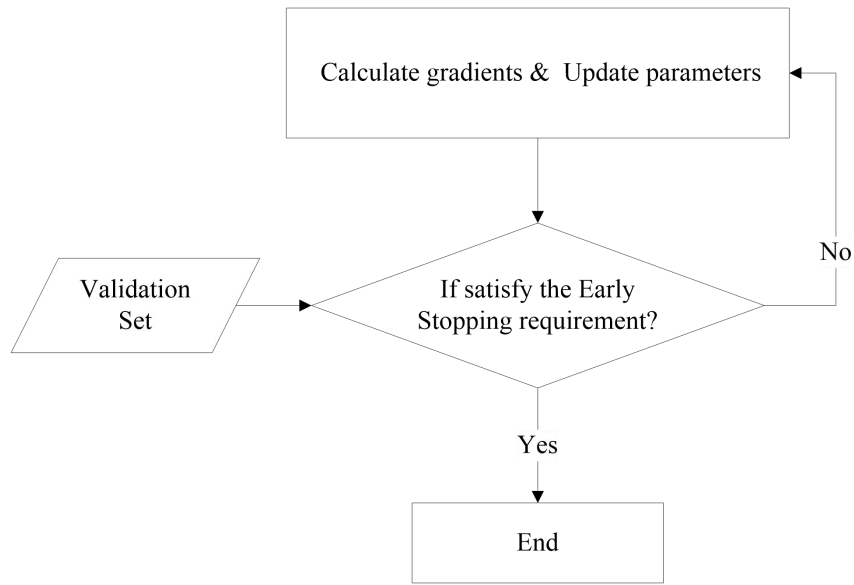

**Fig. 4.** Early Stopping method

## Data preparation and software

### *Data source*

The data source in our study is the market data of constituent stocks in Shenzhen Stock Exchange Component Index (SZSE COMP), which was first released on January 23rd, 1995 and it has been one of the oldest, most complete and influential constituent stock indexes in China's stock market. Our experiment dataset is the daily returns of SZSE COMP constituent stocks from January 1st, 2013 to July 12th, 2017.

The number of constituent stocks of the SZSE COMP index was initially 40 and after May 20th, 2015, expanded from 40 to 500. Since the time horizon studied here is from January 1st, 2013 to July 12th, 2017, we choose the 40 constituent stocks of SZSE COMP released at the beginning of the sample period in order to maintain data consistency. Then we exclude the two stocks which withdrew from the Shenzhen Stock Exchange in 2013 and 2015 respectively. Hence, we have 38 stocks listed in SZSE COMP on December 31st, 2012 as our research dataset.

**Table 1** Statistics of SZSE COMP constituent stocks

| stock code | mean | deviation | Skew | kurtosis | stock code | mean | deviation | Skew | kurtosis |
|---|---|---|---|---|---|---|---|---|---|
| 000001.SZ | 0.00165 | 0.02094 | 0.79647 | 4.33754 | 000709.SZ | 0.00215 | 0.02378 | 0.39903 | 5.97662 |
| 000002.SZ | 0.00189 | 0.02321 | 0.77426 | 3.09714 | 000758.SZ | 0.00063 | 0.02865 | -0.16809 | 3.20825 |
| 000012.SZ | 0.00067 | 0.02670 | -0.21285 | 3.20743 | 000768.SZ | 0.00072 | 0.03100 | -0.12528 | 4.66880 |
| 000039.SZ | 0.00064 | 0.02646 | 0.00905 | 6.42250 | 000776.SZ | 0.00047 | 0.02467 | 0.33241 | 4.12544 |
| 000060.SZ | 0.00071 | 0.02872 | -0.35861 | 5.04713 | 000783.SZ | 0.00158 | 0.02731 | 0.30455 | 3.56617 |
| 000063.SZ | -0.00008 | 0.02491 | -0.17320 | 2.74071 | 000792.SZ | 0.00064 | 0.02518 | -0.47578 | 6.59227 |
| 000069.SZ | -0.00007 | 0.02300 | -0.30535 | 3.86510 | 000858.SZ | 0.00133 | 0.01991 | 0.33430 | 3.22685 |
| 000157.SZ | -0.00052 | 0.02130 | 0.07680 | 5.11125 | 000869.SZ | 0.00071 | 0.02154 | 0.25650 | 2.16371 |
| 000338.SZ | -0.00026 | 0.02240 | -0.23590 | 3.42540 | 000878.SZ | 0.00190 | 0.02594 | 0.30785 | 3.21710 |
| 000401.SZ | 0.00103 | 0.03066 | -0.25465 | 2.51716 | 000895.SZ | 0.00064 | 0.02034 | -0.02842 | 2.74900 |
| 000402.SZ | 0.00108 | 0.02532 | 0.00702 | 3.24280 | 000898.SZ | 0.00163 | 0.02372 | 0.24810 | 4.76655 |
| 000423.SZ | 0.00073 | 0.02037 | -0.00509 | 2.38794 | 000933.SZ | 0.00208 | 0.02720 | 0.12326 | 2.89518 |
| 000425.SZ | 0.00028 | 0.02404 | -0.25180 | 5.65626 | 000937.SZ | 0.00105 | 0.02749 | -0.01482 | 2.27350 |
| 000538.SZ | 0.00192 | 0.02049 | 0.54053 | 2.99083 | 000960.SZ | -0.00043 | 0.02808 | -0.78093 | 4.81746 |
| 000568.SZ | 0.00171 | 0.02167 | -0.00826 | 2.65702 | 000983.SZ | 0.00076 | 0.02614 | 0.11670 | 2.87057 |
| 000623.SZ | 0.00049 | 0.02390 | -0.00311 | 3.70517 | 002024.SZ | 0.00057 | 0.02789 | 0.02627 | 2.47367 |
| 000629.SZ | 0.00141 | 0.02713 | 0.07470 | 3.53965 | 002142.SZ | 0.00208 | 0.02129 | 0.37711 | 3.80465 |
| 000630.SZ | 0.00049 | 0.02039 | 0.38544 | 5.30041 | 002202.SZ | 0.00029 | 0.02539 | 0.08942 | 2.40485 |
| 000651.SZ | 0.00106 | 0.02259 | 0.24542 | 2.53911 | 002304.SZ | 0.00197 | 0.02339 | 0.40102 | 2.17459 |

**Note:** Selected company names of these stocks are listed in Appendix Table 7.

We can see from Table 1 that there are significant differences in terms of statistical characteristics like kurtosis between different stocks and the differences indicate the different underlying mechanisms.

### *Data preprocessing*

The raw data used in this paper is the daily stock returns of 38 constituent stocks listed in Table 1 from SZSE COMP. Assuming that the first transaction price traded during the continuous auction process on $t\text{-}th$ day is denoted as $P_{i,t,f}$ and the last is denoted as $P_{i,t,l}$, we calculate logarithmic stock daily return of the $i\text{-}th$ stock on $t\text{-}th$ day by the following formula:

$$R_{i,t} = \ln P_{i,t,l} - \ln P_{i,t,f} \tag{6}$$

We specify the feature vector to be 20 lagged terms of logarithmic return and their sign directions while setting the task label to be the stock return one period ahead. In order to facilitate the training process, we firstly normalize the return sequence $R_{i,t}$ for each stock by Min-Max method:

$$r_{i,t} = (R_{i,t} - R_i^{mean}) / R_i^{max} \tag{7}$$

Hence, the feature vectors and labels of the task can be denoted as $\{r_{i,t-20}, d_{i,t-20}, r_{i,t-19}, d_{i,t-19}, r_{i,t-18}, d_{i,t-18} \ldots r_{i,t-1}, d_{i,t-1}\}$ and $r_{i,t}$ respectively.

### *Generation of training sets and test sets*

As mentioned above, the sample period used in this study is from January 1, 2013 to July 12, 2017. We select the last 300 days as the test set, while the remaining data is divided into training set and validation set according to the ratio of 7:3.

### *Software*

Table 2 lists the software applied in our experiment. Data preparation is processed by Python 3.6 with the Numpy toolkit and Pandas toolkit. The GP-LSTM unit in our framework is built based on the keras-gp package which extends Keras with GP layers. The GS algorithm is also coded in Python 3.6. Besides, GARCH family models are employed as the benchmarks through the Rugarch package based on R language.

**Table 2** Tasks and corresponding Software

| Task | Software |
|---|---|
| Data Preparing and Handling | Numpy and pandas (Pyhton packages) |
| Grid Search | Python3.6 |
| Deep kernel Gaussian Process | Keras-gp package |
| GARCH Family Models | rugarch(R package) |

**Strategy Design**

In this part, we first calculate the Sharpe Ratio by forecasting the stock conditional return and volatility of next day according to Eq. (2) and then achieve the Sharpe Ratio by the following formula:

$$SharpeRatio_t = Excess\ return / Deviation = (E[y_t] - r^f)/\sigma[y_t] \qquad (8)$$

Referring to the research of Huck et al. (2009) and Thomas Fischer et al. (2017), we go long the top $k$ and short the flop $k$ stocks according to the Sharpe Ratio ranking. The daily return $R^k$ of the long-short strategy can be calculated by Eq. (9). The performance of the constructed portfolio consisting of $2k$ stocks will be examined so that we can jointly evaluate the forecasting performance of stock returns and volatility.

$$R^k = \frac{\sum_{t=1}^{T}\sum_{i=1}^{k}(r_{i,t}^{Long} + r_{i,t}^{Short})}{T} \qquad (9)$$

**Benchmark Models**

Similar to GP model, GARCH family models also can be divided into two parts: the mean equation which depicts the movement of conditional mean, and the volatility equation that depicts the conditional variance of the error term in mean equation. Several commonly used GARCH family models are described in detail as follows.

*Standard GARCH model*

As mentioned previously, in standard GARCH model, the variance of the error term denoted by $h_t$ is impacted not only by the lagged squared error terms $\varepsilon^2_{t-i}$, but also by $h_{t-i}$, the lagged terms of itself. The equation for the GARCH *(p, q)* model is shown below.

$$y_t = x_t^T \beta + \varepsilon_t, \varepsilon_t = \sqrt{h_t} \cdot v_t,$$
$$E(v_t) = 0, D(v_t) = 1, \quad (10)$$
$$h_t = k_0 + \sum_{i=1}^{p} \rho_i h_{t-i} + \sum_{i=1}^{q} \alpha_i \varepsilon^2_{t-i}.$$

### EGARCH model

To reflect the asymmetrical shock, Nelson (1991) proposed the exponential GARCH model (EGARCH) model, which can be written as:

$$y_t = x_t^T \beta + \varepsilon_t, \varepsilon_t = \sqrt{h_t} \cdot v_t,$$
$$E(v_t) = 0, D(v_t) = 1, \quad (11)$$
$$\ln(h_t) = \alpha_0 + \sum_{j=1}^{q}(\alpha_j v_{t-j} - \gamma_j(|v_{t-j}| - E|v_{t-j}|)) + \sum_{j=1}^{p} \beta_j h_{t-j}.$$

where $\alpha_j$ reflects the size effect and $\gamma_j$ reflects the sign effect.

### GJR-GARCH model

The asymmetrical shock of the information flow can also be described by GJR-GARCH model proposed by Glosten, Jaganathan, and Runkle (1993). The GJR-GARCH can mathematically be described as follows:

$$y_t = x_t^T \beta + \varepsilon_t, \varepsilon_t = \sqrt{h_t} \cdot v_t,$$
$$E(v_t) = 0, D(v_t) = 1, \quad (12)$$
$$h_t = k_0 + \sum_{i=1}^{p} \rho_i h_{t-i} + \sum_{i=1}^{q} \alpha_i \varepsilon^2_{t-i} + \gamma \varepsilon^2_{t-1} d_{t-1}.$$

If leverage effect parameter $\gamma \neq 0$, the influence of the information shock is asymmetrical.

In this paper, we choose benchmark models to be SGARCH model, EGARCH model and GJR-GARCH model. The error term of each of the three models is set to be Gaussian distribution, *t* distribution and skewed *t* distribution respectively. The resulting 9 benchmarks are represented as **SGARCH-norm, SGARCH-std, SGARCH-sstd, EGARCH-norm, EGARCH-std, EGARCH-sstd, GJR-GARCH-norm, GJR-GARCH-std, GJR-GARCH-sstd** respectively. In addition, we set $p=1$, $q=1$ (see Eq. (10-12)) for these models because this setting is the most widely used in realistic situations.

# Results

In this section, we focus on presenting experiment findings and start with the exhibition of the overall performance of the constructed portfolios during 2016/04/20 - 2017/07/12. For simplicity, we do not take the transaction fees into consideration and set risk free return to be 0. Then, detailed portfolio performance when $k=15$ reflects that GP-LSTM method can jointly forecast the conditional returns and volatility accurately. Besides, we compare the forecasting accuracy of different models from other perspectives, such as return prediction *Accuracy* and *VaR Kupiec test*.

## Overall portfolio performance

Table 3 in Appendix summarizes the trading strategy performances based on each model. As symbol $k$ denoted the size of portfolio with long position or short position, for $k=3$ or $k=15$, GP-LSTM method is supreme to all GARCH family models in terms of the Sharpe Ratio. In the case of $k=10$, GP-LSTM is still the second best. The high Sharpe Ratio of GP-LSTM implies the strong prediction ability of this method in both conditional returns and volatility.

With regard to metrics of the portfolio daily returns, the performance ranking of GP-LSTM is almost identical to the performance of Sharpe Ratio. The portfolio based on GP-LSTM also provides a low volatility except when $k=10$. The outcome in the two metrics proves the previous conclusion derived by Sharpe Ratio ranking.

## Detailed portfolio performance when $k=15$

As shown in Table 4, GP-LSTM model achieves high daily returns on both long position and short position, which results in the highest hedge return among all the models. This indicates that the new framework can effectively capture the characteristic patterns to distinguish the high-return stocks and low-return stocks.

Value at Risk ($VaR$), the estimation of the maximum potential loss under a given confidence level. Darrel et al. (1997) have made a comprehensive review about $VaR$ theory, and point out that this method can effectively measure the financial market risk. From Table 4, we can also find that GP-LSTM model is superior to all benchmarks in the $VaR$ at 5%,

7.5%, 10% confidence level respectively. Fig. 5 plots the accumulative return of each model on the test period with the black and bold one representing the GP-LSTM based strategy.

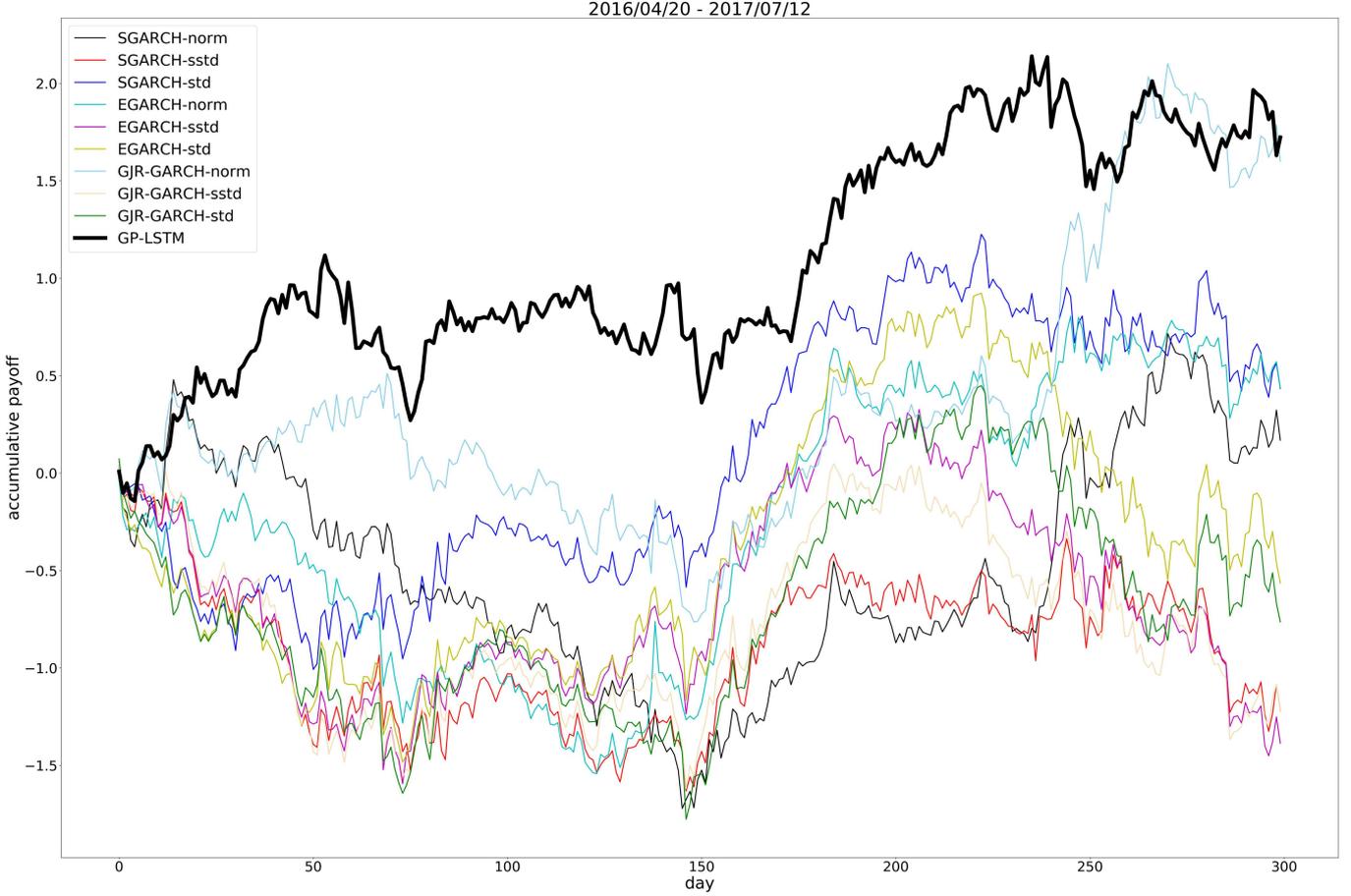

**Fig. 5** accumulative return of portfolio based on each model ( $k=15$ )

**Predictive accuracy**

*Mean Squared Error and Accuracy*

In order to measure the performance of daily return forecast more directly, we calculate the Mean Squared Error ( *Mse* ) and *Accuracy* by the following formulas:

$$MSE = 1/N \sum_{i=1}^{N} (r_i^{forecast} - r_i^{real})^2$$

$$Accuracy = 1/N \sum_{i=1}^{N} I_i, I_i = \begin{cases} 1, & \text{if the forecast direction is right}; \\ 0, & \text{otherwise} \end{cases} \quad (11)$$

A smaller *Mse* or a higher Accuracy means that we have a more accurate forecast of the daily return. For a certain model, we also define a kind of stocks termed *occupying stocks*, whose definition is the stocks on which the model can achieve the best performance in a certain metric. We count the dominant stocks of each model in *Mse* and *Accuracy*, and make the corresponding ranking.

**Table 5** Prediction accuracy of each algorithm

| Model | *Mse* | | | | *Accuracy* | | | |
|---|---|---|---|---|---|---|---|---|
| | *Mse* Value | rank | occupying stocks | rank | *Accuracy* Value | rank | occupying stocks | rank |
| SGARCH-norm | 0.0003497 | 7 | 4 | 4 | 0.5008283 | 7 | 8 | 2 |
| SGARCH-std | 0.0003492 | 6 | 4 | 4 | 0.5026689 | 3 | 4 | 6 |
| SGARCH-sstd | 0.0003489 | 2 | 1 | 9 | 0.5022087 | 5 | 1 | 10 |
| EGARCH-norm | 0.0003489 | 3 | 8 | 1 | 0.4993558 | 9 | 5 | 4 |
| EGARCH-std | 0.0003497 | 8 | 3 | 6 | 0.5026689 | 3 | 5 | 4 |
| EGARCH-sstd | 0.0003492 | 5 | 2 | 7 | 0.5020247 | 6 | 3 | 9 |
| GJR-GARCH-norm | 0.0003485 | 1 | 8 | 1 | 0.4989877 | 10 | 4 | 6 |
| GJR-GARCH-std | 0.0003502 | 9 | 0 | 10 | 0.5038653 | 2 | 4 | 6 |
| GJR-GARCH-sstd | 0.0003491 | 4 | 2 | 7 | 0.5006442 | 8 | 8 | 2 |
| GP-LSTM | 0.0003568 | 10 | 6 | 3 | 0.5086508 | 1 | 10 | 1 |

We can see from Table 5 that GP-LSTM method outperforms all other models in average *Accuracy* and the number of corresponding occupying stocks. Although the GP-LSTM method is inferior to GARCH family models in *Mse*, it still ranks high in number of dominant stocks.

## *Comparison of the VaR forecast results*

As the daily volatility is not visible, we usually measure the accuracy of volatility by *VaR Kupiec test*, which is implemented according to Empirical Failure Rate. Given a fixed confidence level $\alpha$, the event $Hit_i = I\{r_i \leq VaR_i\}$ should subject to the following $0-1\ distribution$:

$$P\{Hit_i = k\} = \alpha^k (1-\alpha)^{(1-k)},\ k = 0, 1. \tag{12}$$

Correspondingly, $\sum Hit_i$ should subject the Bernoulli distribution.

$$\sum Hit_i \sim Bernoulli(\alpha) \tag{13}$$

Let variable $n$ represents $\sum Hit_i$ and $N$ represents sample size, we construct a hypothesis test with the Null hypothesis to be $H_0: n/N = \alpha$. Hence, we can construct LR statistics as follow:

$$LR = 2\ln[(n/N)^n (1-n/N)^{N-n}] - 2\ln[(1-\alpha)^{N-n}(\alpha)^n]$$
$$LR \sim \chi^2(1)$$
(14)

The greater the $P$ value, the closer the empirical failure rate $n/N$ to the confidence level, which means a higher predictive accuracy of volatility. Table 6 displays the average $P$ value and number of occupying stocks of each model in the test set of 38 stocks for 5%, 7.5% and 10% confidence level respectively.

We can see from Table 6 that GP-LSTM ranks high in most cases except when *alpha* = 5%. Although EGARCH-sstd is superior to GP-LSTM in terms of $VaR$ test, this model places 6 among the 10 models in metric $Accuracy$ presented in Table 5, and GP-LSTM model places 1 in this metric. Similarly, the GJR-GARCH-norm model ranks first in the overall $Mse$ and the number of occupying stocks in $Mse$, but in terms of direction $Accuracy$ and $VaR$ prediction, GP-LSTM model is still the superior one. Generally speaking, GP-LSTM method provides an accurate forecast for both conditional return and volatility, which results in the significant performance of the constructed portfolio.

## Further discussion

To further analyze the models' performance under different market conditions, this paper splits the backtesting period of 300-day time horizon into two sub periods equally: April 20th, 2016 – November 29th, 2017 and November 30th, 2016 – July 12th, 2017. The comparison and analysis of the 10 models over the two periods are conducted in this part.

### Sub-period performance

It can be observed from Fig. 6 that, the GP-LSTM method exceeds other models by a large margin with respect to accumulative return during the early half period (2016/04/20 - 2016/11/29). During the period of 2016/11/30 - 2017/07/12, GP-LSTM still has a stable performance. Placing 4 among all the models, GP-LSTM model does not produce an evident advantage against other models, which is significantly different from the performance on the early half period. In the nest section, we will give an in-depth analysis about this difference.

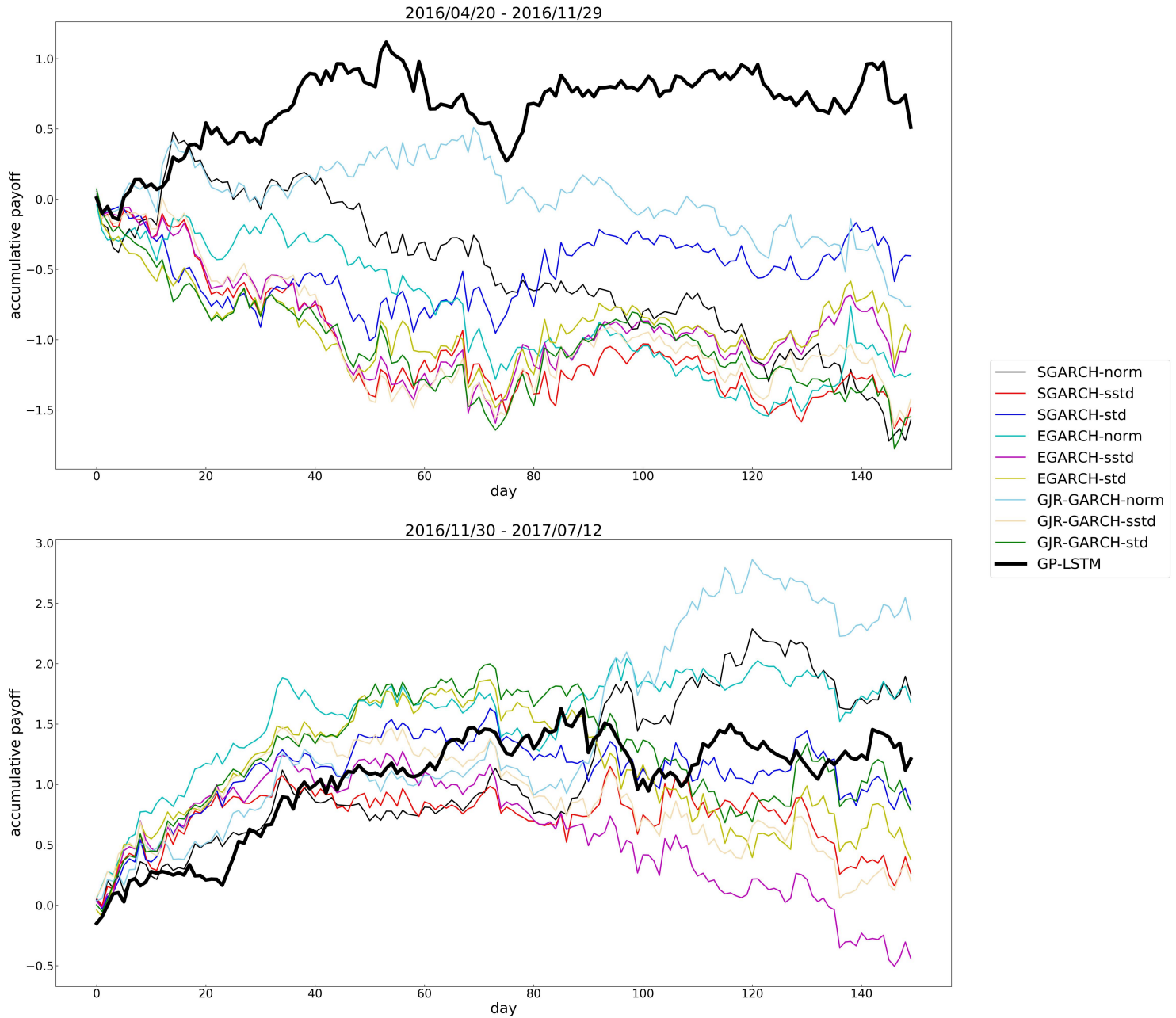

**Fig. 6** sub-period performance of portfolio based on each model ( $k = 15$ )

## Sub-period analysis

For exploring the characteristic of two sub periods mentioned above, we calculate the autocorrelation function ( $acf$ ) and partial correlation function ( $pacf$ ) coefficients for 20 lagged terms. The $acf$ measures the correlation between $y_t$ and $y_{t-k}$ while the $pacf$ measures the partial correlation which removes the effects from other lags. Averaging the

*acfs* and *pacfs* of all the sample stocks on the two sub periods, we plot Fig. 7 and Fig. 8 respectively as follows.

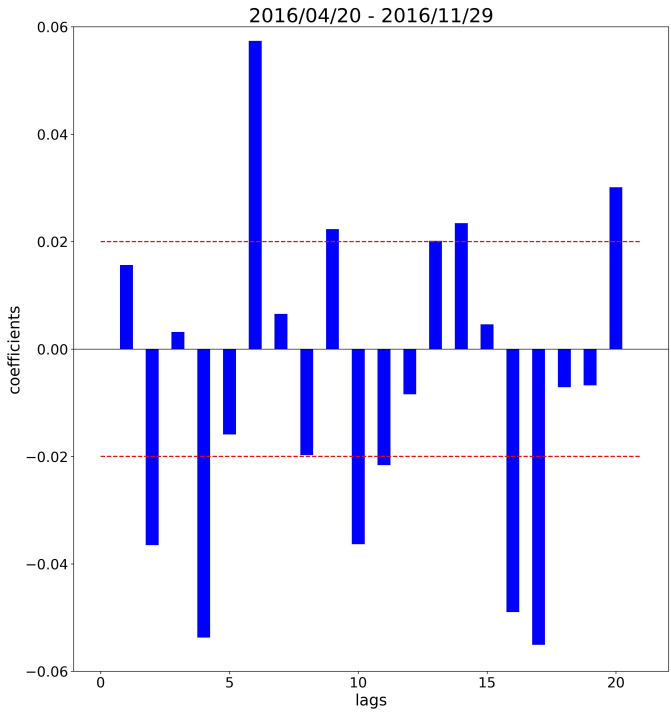
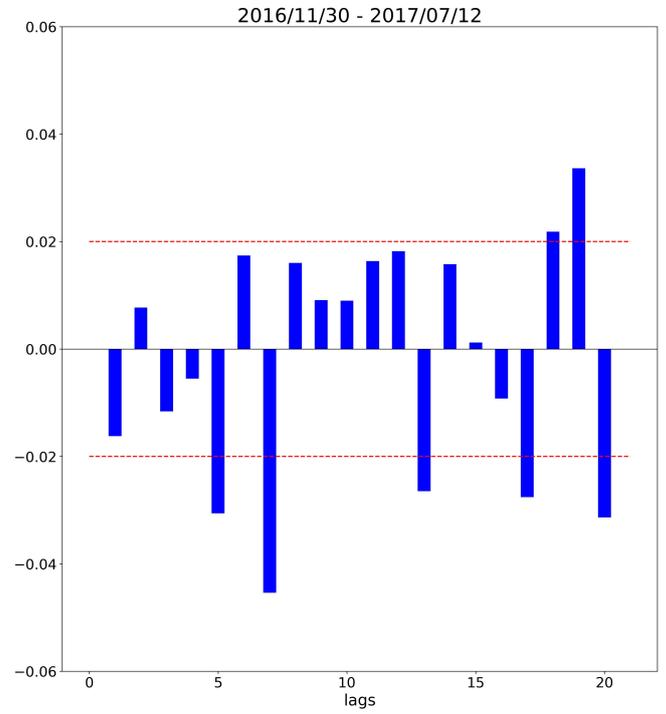

**Fig. 7** averaged *acf* coefficients on the two sub periods

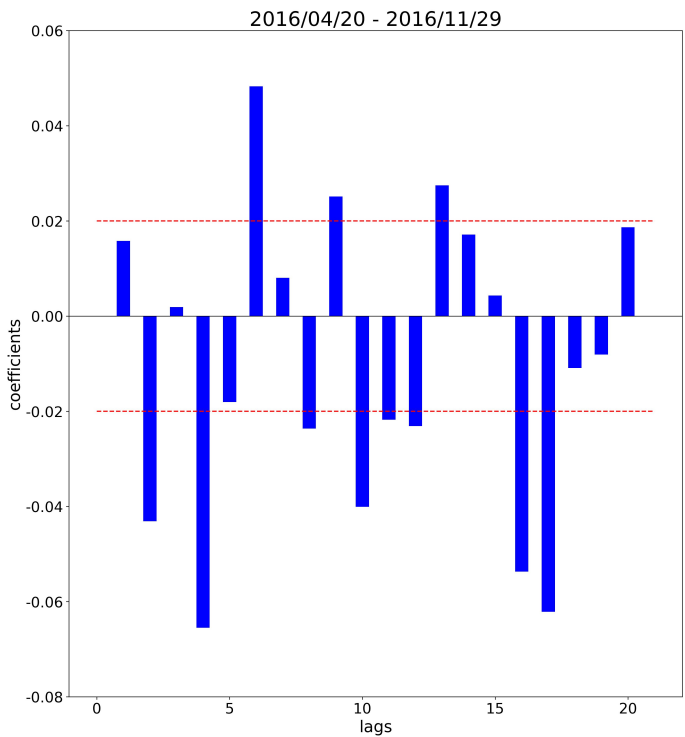
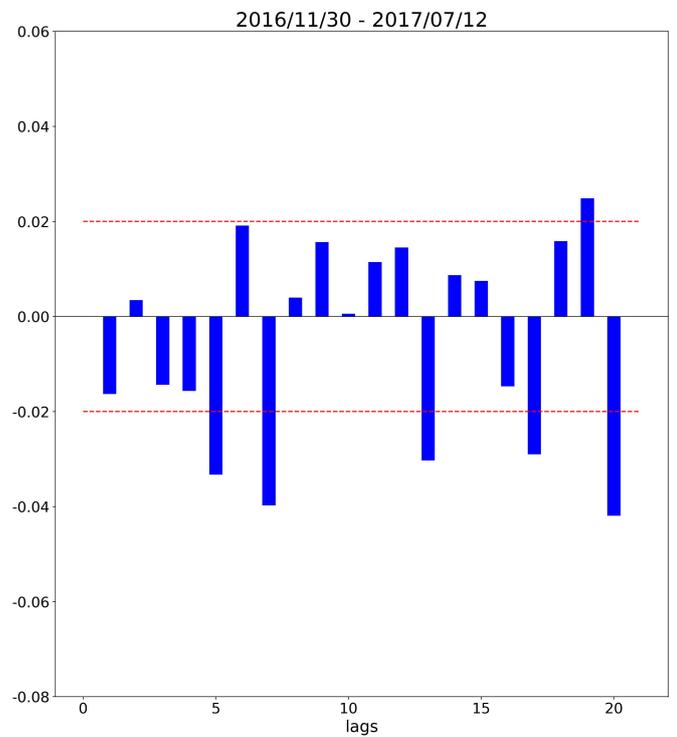

**Fig. 8** averaged *pacf* coefficients on the two sub periods

By plotting the graphs of averaged *acf* and *pacf* coefficients, we can see that the stock data on the first sub period shows a more significant long-term dependency compared with the second sub period. The *K*-line of the SZSE COMP (see Fig. 9) also illustrates that the former sub period is more volatile than the latter. Therefore, we can conclude that GP-LSTM method has an obvious advantage against all other models in the market condition with higher complexity.

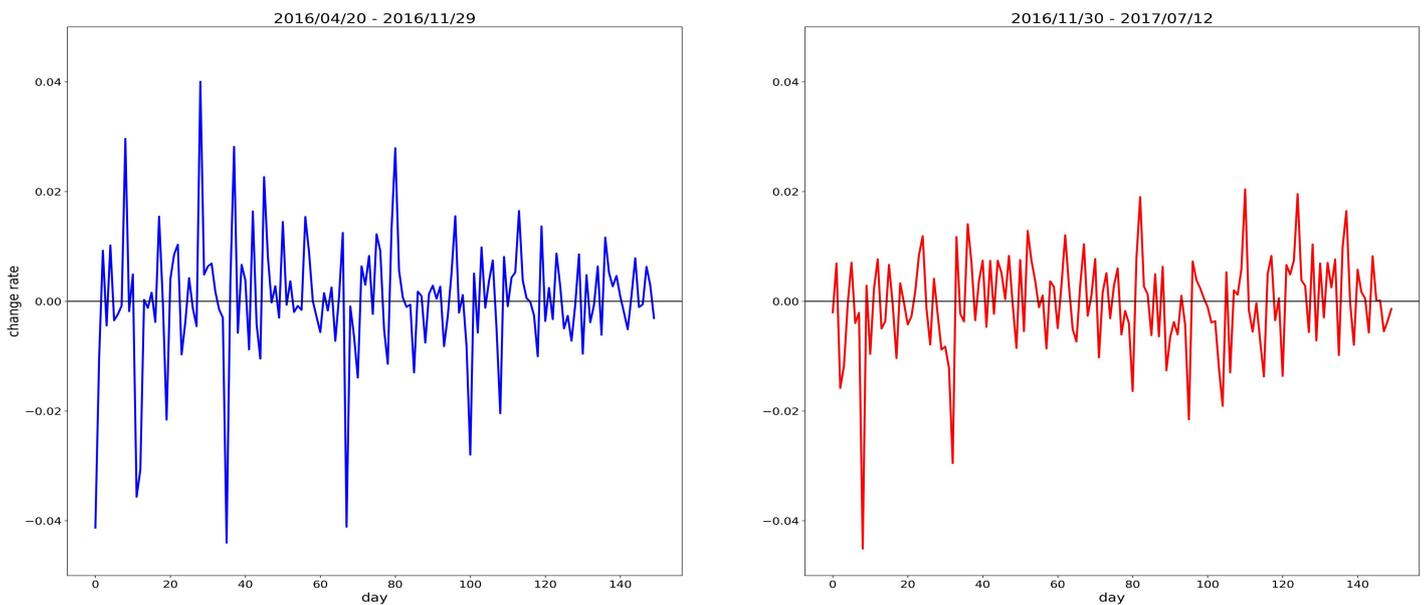

**Fig. 9** Daily return of SZSE COMP on the two sub periods

## Conclusion

In this paper, we discuss the characteristics of previous models for predicting the conditional return and volatility in financial market. It is found that machine learning methods like artificial neural networks (ANN) and support vector regression (SVR) have stronger fitting ability compared with the GARCH family models. However, the ANN and SVR can not be applied to forecast the real volatility directly. Due to the advantage of providing probabilistic output, Gaussian Process (GP) is able to forecast the conditional return and volatility simultaneously. Similar to statistical GARCH family models, the classic GP model still has the limitation of fixed function form. To solve this problem, a deep kernel Gaussian Process (Al-Shedivat et al., 2017) is proposed by extending the basic kernel with deep learning networks. This paper combines GS algorithm with the deep kernel Gaussian Process to forecast the stock returns and volatility. The new method is empirically applied to the dataset covering all constituents of SZSE COMP Index. According to the daily returns and volatility prediction results, we calculate the Sharpe Ratio one period ahead and correspondingly construct the long-short portfolio. It is found that the GP-LSTM method is superior to the benchmark models in terms of the portfolio performance, by which we can jointly evaluate the prediction ability of each model for conditional return and volatility. Further sub-period analysis shows that GP-LSTM method has better generalization ability than the benchmarks in highly volatile and complex periods. Although LSTM networks have been widely used due to the strong fitting ability in modeling sequential data, the input sequence of LSTM is still mapped into a fixed-length vector, which means a loss of information. For future research, we will focus on information compression to promote representation ability for financial time series data and improve the forecasting performance of GP framework.

# Appendix

**Table 3** Daily return and risk of each trading algorithm

| Portfolio Size | Model | Average Daily Return | Standard Deviation | Sharpe Ratio |
|---|---|---|---|---|
| $k=3$ | SGARCH-norm | -0.00089 | 0.03564 | -0.02496 |
| | SGARCH-std | -0.00019 | 0.04090 | -0.00465 |
| | SGARCH-sstd | -0.00056 | 0.04401 | -0.01273 |
| | EGARCH-norm | 0.00047 | 0.03923 | 0.01185 |
| | EGARCH-std | -0.00385 | 0.04261 | -0.09040 |
| | EGARCH-sstd | -0.00250 | 0.04188 | -0.05973 |
| | GJR-GARCH-norm | **0.00191** | 0.03595 | **0.05325** |
| | GJR-GARCH-std | -0.00113 | 0.04495 | -0.02505 |
| | GJR-GARCH-sstd | -0.00010 | 0.04138 | -0.00244 |
| | GP-LSTM | **0.00377** | 0.04033 | **0.09343** |
| $k=10$ | SGARCH-norm | 0.00010 | 0.07049 | 0.00145 |
| | SGARCH-std | 0.00010 | 0.07234 | 0.00142 |
| | SGARCH-sstd | 0.00112 | 0.07799 | 0.01438 |
| | EGARCH-norm | 0.00044 | 0.06760 | 0.00651 |
| | EGARCH-std | -0.00106 | 0.08099 | -0.01313 |
| | EGARCH-sstd | -0.00227 | 0.07260 | -0.03124 |

|  | Model | | | |
|---|---|---|---|---|
|  | GJR-GARCH-norm | **0.00614** | 0.06918 | **0.08870** |
|  | GJR-GARCH-std | -0.00108 | 0.07689 | -0.01402 |
|  | GJR-GARCH-sstd | -0.00134 | 0.07195 | -0.01861 |
|  | GP-LSTM | **0.00327** | 0.06806 | **0.04806** |
| *k=15* | SGARCH-norm | 0.00056 | 0.08560 | 0.00657 |
|  | SGARCH-std | 0.00145 | 0.09112 | 0.01587 |
|  | SGARCH-sstd | -0.00407 | 0.09129 | -0.04457 |
|  | EGARCH-norm | 0.00146 | 0.08152 | 0.01788 |
|  | EGARCH-std | -0.00188 | 0.09625 | -0.01956 |
|  | EGARCH-sstd | -0.00462 | 0.09104 | -0.05077 |
|  | GJR-GARCH-norm | **0.00533** | 0.08844 | **0.06029** |
|  | GJR-GARCH-std | -0.00254 | 0.08823 | -0.02884 |
|  | GJR-GARCH-sstd | -0.00408 | 0.09013 | -0.04527 |
|  | GP-LSTM | **0.00574** | 0.08013 | **0.07167** |

**Table 4** Detailed performance of the daily return from each trading algorithm

| Model | Mean return (long position) | Mean return (short position) | Median |
|---|---|---|---|
| SGARCH-norm | 0.016817 | -0.016254 | -0.0031 |
| SGARCH-std | 0.018046 | -0.016599 | 0.005762 |
| SGARCH-sstd | 0.014373 | -0.018442 | -0.0089 |
| EGARCH-norm | 0.018294 | -0.016837 | 0.000555 |
| EGARCH-std | 0.015527 | -0.017409 | -0.00243 |
| EGARCH-sstd | 0.013907 | -0.018528 | -0.0026 |
| GJR-GARCH-norm | 0.019588 | -0.014256 | -0.00259 |
| GJR-GARCH-std | 0.016097 | -0.018642 | 0.001828 |
| GJR-GARCH-sstd | 0.014068 | -0.018149 | -0.00282 |
| GP-LSTM | 0.017706 | **-0.011963** | 0.003487 |
| Model | 5-percent *VaR* | 7.5-percent *VaR* | 10-percent *VaR* |
| SGARCH-norm | -0.13247 | -0.11439 | -0.09896 |
| SGARCH-std | -0.15042 | -0.1316 | -0.11347 |
| SGARCH-sstd | -0.14164 | -0.1258 | -0.11224 |
| EGARCH-norm | -0.12295 | -0.10431 | -0.09219 |
| EGARCH-std | -0.16004 | -0.13766 | -0.11417 |
| EGARCH-sstd | -0.13530 | -0.12593 | -0.11274 |
| GJR-GARCH-norm | -0.12605 | -0.11135 | -0.09324 |
| GJR-GARCH-std | -0.14419 | -0.12933 | -0.10751 |
| GJR-GARCH-sstd | -0.15204 | -0.12076 | -0.11223 |
| GP-LSTM | **-0.11009** | **-0.09067** | **-0.08059** |

**Table 6** *VaR* test result

| Confidence Level | Model | P value | rank | dominant stocks | rank |
|---|---|---|---|---|---|
| 5.00% | SGARCH-norm | 0.12567 | 10 | 1 | 9 |
| | SGARCH-std | 0.19097 | 7 | 3 | 6 |
| | SGARCH-sstd | 0.22661 | 4 | 8 | 4 |
| | EGARCH-norm | 0.17508 | 8 | 3 | 6 |
| | EGARCH-std | 0.2902 | 2 | 15 | 2 |
| | EGARCH-sstd | 0.32767 | 1 | 18 | 1 |
| | GJR-GARCH-norm | 0.15284 | 9 | 1 | 9 |
| | GJR-GARCH-std | 0.21023 | 5 | 3 | 6 |
| | GJR-GARCH-sstd | 0.22829 | 3 | 5 | 5 |
| | GP-LSTM | 0.2097 | 6 | 12 | 3 |
| 7.50% | SGARCH-norm | 0.06569 | 10 | 1 | 9 |
| | SGARCH-std | 0.15511 | 7 | 5 | 6 |
| | SGARCH-sstd | 0.15518 | 6 | 4 | 7 |
| | EGARCH-norm | 0.08800 | 8 | 1 | 9 |
| | EGARCH-std | 0.21589 | 3 | 10 | 3 |
| | EGARCH-sstd | 0.27236 | 1 | 18 | 1 |
| | GJR-GARCH-norm | 0.07721 | 9 | 2 | 8 |
| | GJR-GARCH-std | 0.16553 | 5 | 6 | 5 |
| | GJR-GARCH-sstd | 0.19095 | 4 | 7 | 4 |
| | GP-LSTM | 0.21769 | 2 | 14 | 2 |
| 10.00% | SGARCH-norm | 0.03263 | 9 | 0 | 7 |
| | SGARCH-std | 0.14274 | 5 | 6 | 4 |
| | SGARCH-sstd | 0.13505 | 7 | 5 | 6 |
| | EGARCH-norm | 0.04682 | 8 | 0 | 7 |
| | EGARCH-std | 0.19584 | 2 | 12 | 2 |
| | EGARCH-sstd | 0.24059 | 1 | 18 | 1 |
| | GJR-GARCH-norm | 0.02804 | 10 | 0 | 7 |
| | GJR-GARCH-std | 0.14307 | 4 | 5 | 6 |
| | GJR-GARCH-sstd | 0.14116 | 6 | 6 | 4 |
| | GP-LSTM | 0.15075 | 3 | 11 | 3 |

**Table 7** Company names of selected constituent stocks from Shenzhen Stock Exchange Component Index

| stock code | Name | stock code | Name |
|---|---|---|---|
| 000001.SZ | Ping An Bank Co., Ltd. | 000709.SZ | HBIS COMPANY LIMITED |
| 000002.SZ | CHINA VANKE CO., LTD. | 000758.SZ | China Nonferrous Metal Industry"s Foreign Engineering and Construction Co., Ltd. |
| 000012.SZ | CSG HOLDING CO.,LTD. | 000768.SZ | AVIC AIRCRAFT Co.,Ltd. |
| 000039.SZ | CHINA INTERNATIONAL MARINE CONTAINERS (GROUP) CO., LTD | 000776.SZ | GF SECURITIES CO.,LTD |
| 000060.SZ | SHENZHEN ZHONGJIN LINGNAN NONFEMET CO.,LTD | 000783.SZ | Changjiang Securities Co., LTD |
| 000063.SZ | ZTE CORPORATION | 000792.SZ | Qinghai Salt Lake Industry Co., Ltd. |
| 000069.SZ | SHENZHEN OVERSEAS CHINESE TOWN CO.,LTD | 000858.SZ | WULIANGYE YIBIN CO.,LTD. |
| 000157.SZ | ZOOMLION HEAVY INDUSTRY SCIENCE AND TECHNOLOGY CO., LTD | 000869.SZ | YANTAI CHANGYU PIONEER WINE COMPANY LIMITED |
| 000338.SZ | Weichai Power Co., Ltd. | 000878.SZ | YUNNAN COPPER CO.,LTD. |
| 000401.SZ | TANGSHAN JIDONG CEMENT CO.,LTD | 000895.SZ | HENAN SHUANGHUI INVESTMENT & DEVELOPMENT CO.,LTD |
| 000402.SZ | FINANCIAL STREET HOLDINGS CO., LTD. | 000898.SZ | ANGANG STEEL CO.,LTD |
| 000423.SZ | Dong-E-E-Jiao Co., Ltd. | 000933.SZ | HENAN SHENHUO COAL & POWER CO.,LTD |
| 000425.SZ | XCMG Construction Machinery Co., Ltd. | 000937.SZ | Jizhong Energy Resources Co., Ltd. |
| 000538.SZ | YUNNAN BAIYAO GROUP CO.,LTD. | 000960.SZ | Yunnan Tin Co., Ltd. |
| 000568.SZ | LUZHOU LAO JIAO CO.,LTD | 000983.SZ | SHANXI XISHAN COAL AND ELECTRICITY POWER CO.,LTD |
| 000623.SZ | Jilin Aodong Pharmaceutical Group Co., Ltd. | 002024.SZ | SUNING.COM CO.,LTD. |
| 000629.SZ | Pangang Group Vanadium Titanium & Resources Co., Ltd. | 002142.SZ | BANK OF NINGBO CO., LTD |
| 000630.SZ | Tongling Nonferrous Metals Group Co.,Ltd. | 002202.SZ | XINJIANG GOLDWIND SCIENCE&TECHNOLOGY CO.,LTD |
| 000651.SZ | GREE ELECTRIC APPLIANCES,INC.OF ZHUHAI CO.,LTD | 002304.SZ | Jiangsu Yanghe Brewery Joint-Stock Co.,Ltd. |

## Abbreviations

GP: Gaussian Process; LSTM: long-short term memory; GP-LSTM: a Gaussian process with the kernel learned by LSTM; GS: grid search algorithm; SZSE COMP: Shenzhen Stock Exchange Component Index; ARCH: Autoregressive conditional heteroskedasticity; GARCH: Generalized Autoregressive Conditional Heteroskedasticity; ANN: Artificial Neural Networks;
SVM: support vector machine; SVR: support vector regression; $RBF$ : Radial Basis Function; $VaR$ : Value at Risk

## Availability of data and materials

The datasets used during the current study are collected from the database of Wind Information Co., Ltd (https://www.wind.com.cn/).

## Competing interests

The authors declare no conflict of interest.

## Funding

This work was supported by National Science Foundation of China, Key Project, #71932008; National Science Foundation of China, Project, #71771204.

## Authors' contributions

YS: Resources, Supervision, Funding acquisition; WD: Formal analysis, Software, Visualization, Writing - original draft; WL: Conceptualization, Methodology, Writing - review & editing, Validation; BL: Data curation, Investigation. All authors read and approved the final manuscript.